# Comments on "Consensus and Cooperation in Networked Multi-Agent Systems" [1]

PAVEL CHEBOTAREV,[2] *Member IEEE*

**Key words**: consensus algorithms, cooperative control, flocking, graph Laplacian, networked multi-agent systems

The objective of this note is to give several comments regarding the paper [1] published in the Proceedings of the IEEE and to mention some closely related results published in 2000 and 2001. I will focus on the graph theoretic results underlying the analysis of consensus in multiagent systems.

As stated in the Introduction of [1], "*Graph Laplacians* and their spectral properties […] are important graph-related matrices that play a crucial role in convergence analysis of consensus and alignment algorithms." In particular, the stability properties of the distributed consensus algorithms

$$\dot{x}_i(t) = \sum_{j \in N_i} a_{ij}(t)\bigl(x_j(t) - x_i(t)\bigr), \quad i = 1,\ldots,n \tag{1}$$

for networked multi-agent systems are completely determined by the location of the Laplacian eigenvalues of the network. The convergence analysis of such systems is based on the following lemma [1, p. 221]:

*Lemma 2*: (spectral localization) Let $G$ be a strongly connected digraph on $n$ nodes. Then $\mathrm{rank}(L) = n - 1$ and all nontrivial eigenvalues of $L$ have positive real parts. Furthermore, suppose $G$ has $c \geq 1$ strongly connected components, then $\mathrm{rank}(L) = n - c$.

Here, $L$ is the Laplacian matrix of $G$, i.e., $L = D - A$, where $A$ is the adjacency matrix of $G$, and $D$ is the diagonal matrix of vertex out-degrees.

I would like to make four comments regarding this lemma.

**1.** The last statement of the lemma is wrong. Indeed, recall that the strongly connected components (SCC's) of a digraph $G$ are its maximal strongly connected subgraphs. Let, for example, $G$ be a converging tree (in-arborescence), i.e., a directed tree having a node $r$ (a root) such that all nodes can be linked to $r$ via directed paths (for $r$ itself it is a path of length 0). Then $G$ has $c = n$ strongly connected components, so Lemma 2 yields that $\mathrm{rank}(L) = n - c = 0$. But in fact, $\mathrm{rank}(L) = n - 1$. To make the last statement of Lemma 2 valid, one should additionally require that all the SCC's of $G$ are disjoint.

**2.** In [1], the proof of the rank property (the first statement of Lemma 2) is attributed to [3]. Let me note that the general problem of finding $\mathrm{rank}(L)$ for digraphs was solved earlier in [2]. More specifically, by Proposition 11 of [2], for any digraph $G$, $\mathrm{rank}(L) = n - d$, where $d$ is the so-called *in-forest dimension* of $G$, i.e., the minimum possible number of converging trees in a spanning converging forest of $G$. Furthermore, it was shown (Proposition 6) that the in-forest dimension of $G$ is equal to the number of its sink SCC's (the SCC's having no edges directed outwards) and that the in-forest dimension of a strongly connected digraph is one (Proposition 7)[3]. A corrected version of the above Lemma 2 immediately follows as a special case.

**3.** Remark 1 given after Lemma 2 says: "Lemma 2 holds under a weaker condition of existence of a directed spanning tree for $G$." Here, by Lemma 2 the authors presumably mean the conclusion that $\mathrm{rank}(L) = n - 1$ and by a directed tree they mean a converging tree. Next, they note that such a weaker condition has appeared in several papers published in 2003 and 2005. Let us observe that the existence of a spanning converging tree for $G$ is tantamount to $d = 1$, so this statement follows from Proposition 11 of [2].

---

[3] These results have also been presented in [4].

**4.** For the study of alignment algorithms for arbitrary digraphs, it is important to observe that the statement of Lemma 2 that "all nontrivial eigenvalues of $L$ have positive real parts" holds true for any digraphs [5, Proposition 9], and not only for strongly connected digraphs or digraphs with spanning converging trees.

In Section II.C of [1], a discrete-time counterpart of the consensus algorithm (1) is considered

$$x_i(k+1) = x_i(k) + \varepsilon \sum_{j=1}^{n} a_{ij}(x_j(k) - x_i(k)), \quad i = 1,\ldots,n, \qquad (2)$$

where $\varepsilon > 0$ is the step size. In the matrix form, (2) is represented as follows:

$$x(k+1) = Px(k), \qquad (3)$$

where $P = I - \varepsilon L$ is referred to in [1] as the *Perron matrix with parameter* $\varepsilon$ of $G$.

The matrices $P = I - \varepsilon L$ were studied in [2] and [5]; in particular, (i) of Lemma 3 in [1] actually coincides with Proposition 12 of [2].

Finally, let me mention a few additional results [2, 5] that are applicable to the analysis of consensus algorithms (1) and (3) and flocking algorithms. In the general case where the primitivity of a stochastic matrix $P$ is not guaranteed and the sequence $P, P^2, P^3,\ldots$ may diverge, the *long-run transition matrix* $P^\infty = \lim_{m \to \infty} m^{-1} \sum_{k=1}^{m} P^k$ is considered. $P^\infty$ always exists and, by the *Markov chain tree theorem* [6, 7], it coincides with the *normalized matrix* $\bar{J}$ *of maximal in-forests of* $G$. $\bar{J}$ is the eigenprojector of $L$; by Proposition 11 of [2], $\mathrm{rank}(\bar{J}) = d$, where $d$ is the in-forest dimension of $G$. The columns of $\bar{J}$ span the kernel (null space) of $L$; as a result, they determine the main properties of the trajectories of (1) and the flocking trajectories [8] in the general case. The elements of $\bar{J}$ were characterized in graph theoretic terms in Theorems 2' and 3 of [2]; a finite algebraic method for calculating $\bar{J}$ was proposed in [5] (see also [4]).

Thus, [2, 5, 4] published before the recent avalanche of papers on distributed consensus algorithms ([2] and [5] were sent to J.A. Fax in 2001 and a reference to [4] was sent to R. Olfati-Saber apropos of Lemma 2 in 2003, both on their requests) contained the basic graph theoretic results needed for the analysis of these algorithms. A number of related theorems were proved in [9] and [10]. Some of these results were surveyed in [11].